# Water Electrolysis and Energy Harvesting with 0D Ion-Sensitive Field-Effect Transistors

N. Clément[1], K. Nishiguchi[2], J.F. Dufreche[3], D. Guerin[1], A. Fujiwara[2], & D. Vuillaume[1]

(1) Institute of Electronics, Microelectronics and Nanotechnology, CNRS, Avenue Poincaré, 59652, Villeneuve d'Ascq France

(2) NTT Basic Research Laboratories, 3-1, Morinosato Wakamiyia, Atsugi-shi, 243-0198 Japan

(3) Université de Montpellier 2, Institute of separation chemistry Marcoule (ICSM UMR CEA/CNRS/UM2/ENSCM), 30207 Bagnols sur Cèze cedex France

**The relationship of the gas bubble size to the size distribution critically influences the effectiveness of electrochemical processes. Several optical and acoustical techniques have been used to characterize the size and emission frequency of bubbles. Here, we used zero-dimensional (0D) ion-sensitive field-effect transistors (ISFETs) buried under a microbath to detect the emission of individual bubbles electrically and to generate statistics on the bubble emission time. The bubble size was evaluated via a simple model of the electrolytic current. We suggest that energy lost during water electrolysis could be used to generate electric pulses at an optimal efficiency with an array of 0D ISFETs.**

Water electrolysis is one of the most promising "green" approaches for producing hydrogen ($H_2$) as a fuel source.[1] As a key issue in the operation of the International Space Station and the mission to Mars,[2] water electrolysis is utilized as a regenerative life support system and is part of the energy conversion system. The behavior of gas bubbles in water electrolysis is a typical interfacial phenomenon. Many experiments, including optical,[3,4] acoustical,[5,6] and other approaches (e.g., electrical impedance),[7] have been performed to determine the bubble size distribution in samples of water or blood. Here, we propose a new technique based on a "zero-dimensional" (0D) ion-sensitive field-effect transistor (ISFET), which transduces the energy lost during bubble emission in electric pulses. The developed system does not require light, is sensitive to a single bubble, and has no intrinsic limitation in bubble size. Figure 1 shows a schematic view of the device, which is composed of a platinum (Pt) microelectrode and 0D ISFETs, in 3D view (a) and side view (b).

A 0.2-µL droplet of water containing NaCl was placed in a microbath,[8] which was delimited by a SU8 polymerized resist (Shipley) to isolate the water microbath from the outside connection pads (see supplementary information for fabrication details). We studied water electrolysis in the presence of a salt (NaCl) to increase the electrical conductivity of the electrolyte. Salt is often used to improve $H_2$ emission efficiency,[9] through the following well-known electrochemical process:

at the anode: $2Cl^- \rightarrow Cl_2 + 2e^-$

at the cathode: $2H_2O + 2e^- \rightarrow H_2 + 2OH^-$ (1)

The Pt microelectrode was used to bias the electrolyte at voltage $V_{ANODE}$. From an electrochemical perspective, the Pt electrode was the anode at which the $Cl_2$ bubbles appeared (see optical microscopic image in Fig. 1c). The cathode (grounded) was a highly doped silicon (Si) lead at which the small $H_2$ bubbles could be distinguished (Fig. 1c and movie in Supplementary Information). Si was preferred to Pt for this proof of principle because, for Pt cathodes, two or three nucleation points were usually observed, even for sub-micrometric electrodes. $Cl_2$ had a strong affinity to Pt, such that the $Cl_2$ bubbles were very large compared to those of $H_2$ (see Fig. 1c). Consequently, the emission frequency of $Cl_2$ bubbles was so small that we could study $H_2$ bubble emission alone over a few minutes.

We used the 0D ISFET (Fig. 1d) to convert the electrochemical energy lost at the cathode interface (i.e., the fluctuation of the Debye layer of ions at each bubble emission) to a useful electrical signal. We observed a spike in the output current of the ISFET at each nucleation / generation of a $H_2$ bubble (Fig. 1d and Fig. 3 for real signals). The amplitude and frequency of these spikes were controlled by the ion concentration and applied $V_{ANODE}$ (see below). The diameter and length of the 0D ISFET were ~25 nm, according to scanning electron microscopy (SEM) (See Supplementary Information for fabrication details).

We used the term "0D" to distinguish our device—a punctual undoped sensor—from doped nanowire (NW) sensors, whose length is usually in the micrometer range.[10-11] This device operates as a single-electron transistor (SET) up to nitrogen temperature[12,13] and as an ultrasensitive FET at room temperature, with subelementary charge sensitivity.[14,15] When placed in contact with a liquid (as in Fig. 1a), the SET does not lose any charge sensitivity (see supplementary information, Fig.S3).[8] Thus, the noise induced by the Brownian motion of ions at equilibrium near the Si quantum dot (40 nm from the $SiO_2$ / liquid interface) is not the limiting source of noise, at least for NaCl ions.[8] When not in equilibrium (e.g., during water electrolysis), a larger noise is observed, which will be detailed below.

The main properties of the 0D ISFET are demonstrated in Fig. 2. We used the system's dual-gate property (liquid gate on top, and Si gate on back; see Fig. 1b) to tune the drain current-anode voltage curves to the preferred range with the back-gate voltage (Fig. 2a).[16,17] The gate leakage current between the channel and electrolyte was negligible (<< pA). The sensitivity to the ionic strength was close to that of a full Nernstian response (~60 mV/pNa) at constant pH,[8] with some dispersion (Fig. 2c) due to the difficulty of precisely controlling the ion concentration in a microdroplet. A similar response has been reproduced with a microfluidic channel.[18] This response is very different from that of 1D ISFETs that show less ionic sensitivity.[19,20] This finding suggests that 0D ISFETs with only a few Si-OH, Si-O$^-$, and Si-OH$_2^+$ groups at the surface behave differently from 1D[10-13,21,22] and 2D ISFETs.[21]

At sufficiently large $V_{ANODE}$ (typically > 3 V), we observed the emission of $H_2$ bubbles (see video in Supplementary Information), which mainly appeared at an electrolytic current of 1 µA in our microbath ([NaCl] = 100 µM/L). Figure 3a shows the *current-time (I-t)* electrical characteristics measured at the same time in two 0D ISFETs located 500 µm from each other in the same microbath. We periodically observed an abrupt drop in current in both transistors, followed by a progressive recovery. Although the current level was different in both 0D ISFETs (due to device dispersion), the corresponding variations in the input liquid-gate voltage $\Delta V_{ANODE} = \Delta I/g_m$, where $\Delta I$ is the current variation and $g_m$ the transconductance $(\partial I/\partial V_{ANODE})$, were superimposed for both devices (Fig. 3b, see Supplementary Information, Fig. S4 for $g_m$ extraction). Thus, if each spike corresponded to an emitted $H_2$ bubble, then a change in the electrochemical potential of the droplet at this stage could be reproducibly measured with both sensors.

We were able to tune the frequency and amplitude of the spikes with the $V_{ANODE}$ and NaCl concentration. Figure 4a shows the $\Delta V_{ANODE}$ extracted from the current of the 0D ISFETs at different $V_{ANODE}$ values. At 3 V (i.e., just below the threshold voltage for bubble generation), the 0D ISFET registered a large voltage drift (~20 mV). At 3.5 V, bubbles and peaks were observed, as described previously (Fig. 3). At 4 V, the amplitude and frequency of the peaks were increased. In the last case, the signal was not perfectly periodic, due to fluctuations in the peak amplitude. Increasing the NaCl concentration to 1 M/L enabled a good control of the shape of the peaks (Fig. 4b). In that case, the bubbles were observed at $V_{ANODE} > 4.5$ V.

The bubble emission frequency also increased with increasing $V_{ANODE}$. By taking the Fast Fourier Transform (FFT) of this curve (see Supplementary Information, Fig. S5a), we easily obtained the emission frequency $f_e$ from the first FFT peak. In contrast, no current peak was directly observed from the FFT of the electrolysis anode current (see Supplementary Information, Fig. S5b). The electrolysis current measured by an ampermeter was fitted by $\Delta Q \cdot f_e$ (Fig. 4c), where the constant $\Delta Q \approx 0.92$ µC corresponds to the total amount of electrons (~5.8 x $10^{12}$) required to emit 1 $H_2$ bubble. Considering that 2 electrons were needed to emit 1 $H_2$ molecule (see Eq. 1), we deduced that 1 $H_2$ bubble was composed of 4.8 pM $H_2$, corresponding to a bubble radius $r \approx 21$ µm with a gas evolution efficiency $f_G = 1$. The gas evolution efficiency, $f_G$, accounts for the fact that only a fraction of the total hydrogen formed is transferred into the gas bubble. $f_G$ can be less than 1[4,22] but because of the 1/3 power, the resulting effect on the radius $r$ is weak. The optical microscopic images gave $r$ in the range 18-24 µm (see Supplementary Information, Fig. S6), in good agreement with this estimate.

The above approach can be summarized with Eq. 2:

$$r = \left( \frac{3 V_0 f_G I \Delta t}{4 \pi F \cdot n} \right)^{1/3} \quad (2)$$

where $V_0$ is the molar volume of the gas, $n$ is the number of electrons required for 1 molecule of gas, $\Delta t$ is the time between the emissions of two bubbles, $I$ is the anode current, and $F$ is the Faraday constant. Using Eq. 2, we obtained the bubble radius dispersion (Fig. 4d) from the *I-t* curves (Fig. 4b) which is much smaller (~4%) than typical dispersion (25%) usually observed when many nucleation points coexist.[23]

During bubble emission, an important pH gradient occurs between the 2 electrodes. It can be evaluated from the following argument. At the anode, $Cl_2$ can be dismuted into $Cl^-$ and HClO which is a weak acid. At the cathode, emitted $OH^-$ ions increase the pH. Typically, during 1s lifetime of one bubble, 10 pM of $OH^-$ are emitted. The surrounding volume corresponds to a micrometer around the bubble so that the local concentration is evaluated to be close to 1M/L. Thus, at the cathode pH~14. Because of diffusion and migration, this local concentration is smeared out. For $OH^-$, the diffusion coefficient is $2.1.10^{-9}$ m²/s and typical distance between electrodes being about 500 μm, every second $OH^-$ ions diffuse to 100 μm so that the resulting diffusion time is 25s and the migration time is 0.1s. Thus, after this relatively short time, the surrounding electrolyte is stirred and there is a possibility of stationary state. The ionic concentrations are modified by bubble emission, nevertheless, because of dilution (see Supplementary Information for detailed discussion), these effects are expected to be small, but at low salt concentration. It may be an explication of the larger signal measured at low salt concentration.

We propose two possible mechanisms for the electrical detection of bubbles with our 0D ISFET: 1) the resistance at the cathode interface was decreased due to an increase in the ion concentration when each bubble was detached from the cathode, or 2) the electrode surface increased in a similar manner to what is observed in polarography.[24] In both cases, a similar equivalent circuit was used to model the electrochemistry in the microbath. When there was no electrolytic current, the electrochemical potential was in equilibrium, and the output current was stable (no drift; see Supplementary Information, Fig. S3). When there was an electrolytic current, we considered the simple equivalent circuit shown in Fig. 5a. The Nernst potential $V_N$ of the electrolyte was coupled to both the anode and the cathode through the usual double-layer interfaces, which were modeled as a resistance in parallel with a capacitor. $R_a$, $R_c$, $C_a$, and $C_c$ are the resistances and capacitances at the anode and cathode, respectively.

Considering a sudden and short local decrease of $R_c$ at each emission of a $H_2$ bubble from $R_{c+}$ to $R_{c-}$ (as shown in Fig. 5b), we derived, from the equivalent circuit shown in Fig. 5a, an equation for the bubble growth mechanism:

$$V_N = \left[ \frac{R_{c+}}{R_a + R_{c+}} + \beta e^{-t/\tau_{ac}} \right] V_{ANODE} \text{ with } \tau_{ac} = \frac{R_a R_{c+}}{(R_a + R_{c+})}(C_a + C_c) \text{ and } \beta = \left[ \frac{R_{c+}}{R_a + R_{c+}} - \frac{R_{c-}}{R_a + R_{c-}} \right] \quad (3)$$

where $\Delta V = V_N (t = 0) - V_N (t = \infty) = \beta$. $V_{ANODE}$ is the maximum voltage fluctuation amplitude of $V_N$ (Fig. 5b) and $\tau_{ac}$ is the recovery time. Reasonable fits from zooms of Fig. 4a and 4b were obtained by considering [$\beta \cdot V_{ANODE}$, $\tau_{ac}$] = [80 mV, 1.5 s] at [NaCl] = $10^{-4}$ M (Fig. 5c) and [18 mV, 14 ms] at [NaCl] = 1 M (Fig. 5d). Once the gas bubble was nucleated in the supersaturated layer, it began to generate microconvection at the interface between the gas and the liquid on the electrode surface. When a bubble detached from the electrode surface, it started to move upward, entraining the electrolyte. Macroconvection was introduced by the buoyancy flow of the gas-liquid dispersion. We expected that the ion concentration at the cathode interface would be momentarily increased. The lower the ion concentration, the larger the change, in agreement with the larger $\Delta V_N$ measured at [NaCl] = $10^{-4}$ M compared to [NaCl] = 1 M. The $\tau_{ac}$ value was larger for smaller NaCl concentrations, according to Eq. 3, because the resistances $R_a$ and $R_c$ were drastically increased at small NaCl concentrations.

The other possibility was that the effective surface of the cathode changed in a manner similar to that in polarography. However, this possibility cannot explain the difference in results obtained for different ion concentrations. In addition, if we consider the case of an anode that is progressively covered by $Cl_2$ bubbles without affecting the 0D ISFET average current, we can conclude that a thin electrolyte layer may always remain at the $SiO_2$ surface. The distance of the 0D ISFET to the electrode seems to play a larger role at high ionic strength, because we observed the signal only for the closest transistor that was located 100 µm from the bubble emission. This finding may be related to the smaller signal (~10 mV) detected in that case.

In addition to gathering information on the bubble emission, this system could be used as an efficient pulse generator (DC to pulse convertor). At the single-transistor level, the efficiency was low because the power consumed for electrolysis (~3.5 µW) was much larger than the output power of the pulse signal (~$V_D.\Delta I.\tau_{ac} / \Delta t$ ~ 50 nW). However, pulses were obtained simultaneously for all transistors below the microbath. Considering that two million 0D transistors could fit below the microbath, using large-scale integration and biasing these transistors in the subthreshold region ($\Delta I$ is maximum at a given $\Delta V_N$, so all of the transistor power was used only for pulse generation), we obtained an output pulse power of 500 µW, which led to a pulse power efficiency of ~99% (pulse power vs pulse power + water electrolysis power) in addition to $H_2$ generation. We can also note that the amplitude of the generated signal is on a par with that of the action potential in axone/neuron. We envision the possibility to use this system in a physiological environment (mainly water) as an artificial action potential generator to perform local neuron stimulation with tunable amplitude and frequency.

Such a system is also interesting for water electrolysis application. However, monitoring massive gas generation is not appropriate with such system using a single cell. Such experiment has been performed at high anode voltage (6V) and the resulting signal loses its regularity (see Fig.S7). Taking into account the small size of our water electrolysis microsystem, parallel integration off such systems for larger gas generation may be possible, though. Short term application of our device could be lab-on chip devices with hydrogen emission control and storage.

In conclusion, we have proposed a new technique to study bubbles and gather part of the energy lost during water electrolysis using a 0D ISFET. In this system, the punctual sensors are buried under an electrolytic bath. Using this method, we performed individual $H_2$ bubble counting, evaluated the bubble radius, and generated statistics on the bubble emission time. This method could be useful for experiments performed in the dark, or to evaluate the presence of nanobubbles where CCD cameras are prohibited. Another potential use is to investigate the macroconvection process induced by bubble detachment through local changes in the Nernst potential. Because the full Nernstian response to salts can only be observed for 0D ISFETs, this device promises further unique applications in chemistry, biochemistry, and bioelectronics, in which low-power pulse generators are required.

Acknowledgements

N.C would like to thank Nord Pas de Calais district and Université Lille1 for funding (Project Singlemol)

CAPTIONS.

Fig. 1 (a) 3D schematic view of the device, composed of 0D ISFETs in contact with a water droplet containing NaCl ions and delimited by SU8 resist walls. A Pt electrode was used to bias the droplet and generate an electrolytic current between the anode and cathode (highly doped Si lead). (b) Schematic side view of the experimental setup, showing bubble emission at the anode / cathode and the Si quantum dot, which was buried below 40 nm of thermal $SiO_2$, in contact with the droplet. (c) Optical microscopic image corresponding to the schematic view shown in (a). A 200-µL droplet of water containing NaCl was inserted in the microbath. Small $H_2$ bubbles formed at the cathode. A large $Cl_2$ bubble can be observed at the anode. Bottom: schematic view of the input DC signal ($V_{ANODE}$). (d) SEM image of the 0D ISFET. Scale bar is 20 nm. Bottom: schematic view of the output AC signal measured by the 0D ISFET.

Fig. 2 (a) $I$-$V_{ANODE}$ for different $V_{BG}$ values for a 0D ISFET with a liquid gate at room temperature. Subthreshold swing S ~ 140 mV/dec. (b) Threshold variation shift observed in four 0D ISFETs at different NaCl concentrations at constant pH ~ 6. Full Nernstian sensitivity was generally observed.

Fig. 3 (a) Drain current measured in two 0D ISFETs when $V_{ANODE}$ was sufficiently large to produce bubbles. Spikes can be observed. (b) Variation of electrochemical potential estimated from (a) for both transistors, considering the transconductance $g_m$ according to $\Delta V_{ANODE} = \Delta I_D/g_m$.

Fig. 4 (a) Time-dependent measurements in two 0DISFETs (as in Fig. 3b) at different $V_{ANODE}$ values and [NaCl] = $10^{-4}$ M. (b) Time-dependent measurements in a 0D ISFET at different $V_{ANODE}$ values with [NaCl] = 1 M. (c) Current measured in the anode with a semiconductor signal analyzer and $f_e$.9 x $10^{-7}$. $f_e$ was obtained from the FFT power spectrum of the time-dependent measurements. (d) Measured time $\Delta t$ between two spikes at $V_{CATHODE}$ = 4.8 V for 45 s emission and related statistics (inset). Fit was obtained with a Poisson distribution. (e) Histograms of $\Delta t$ and bubble radius (from Eq. 2) obtained with data shown in Fig. 4c. Bubble radius is consistent with the optical image, showing the correlation between spikes and bubble emission.

Fig. 5 (a) Equivalent circuit considering coupling of the electrolytic potential $V_N$ to both anode and cathode via double-layer capacitances and interfacial resistances. (b) Reasonable fit was obtained for $\Delta V_N$ at two different ionic strengths. In Eq.4 , we supposed that a sudden decrease in resistance $R_{c+}$-$R_{c-}$ occurred at the cathode interface at the moment of bubble detachment.

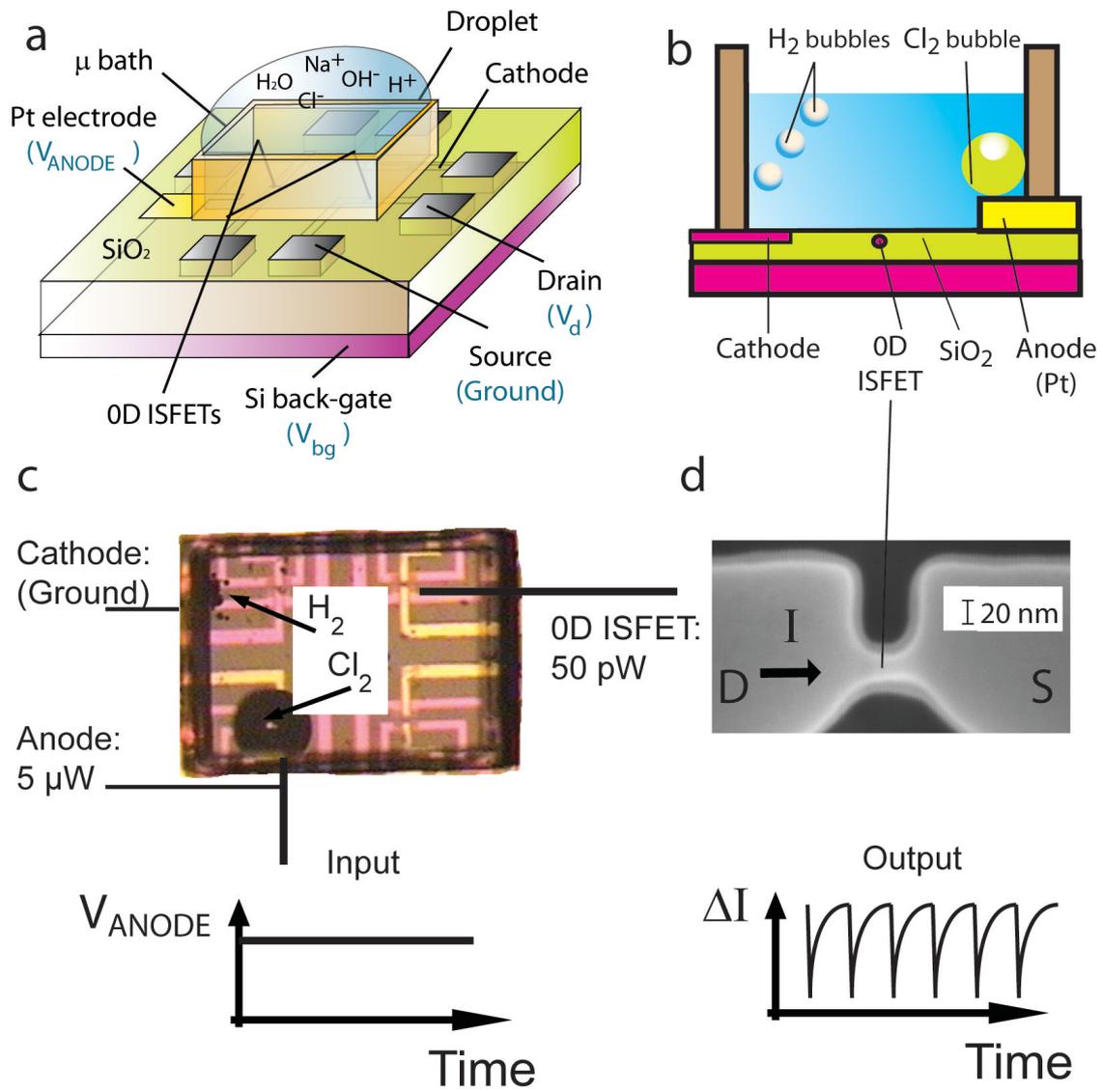

Fig.1

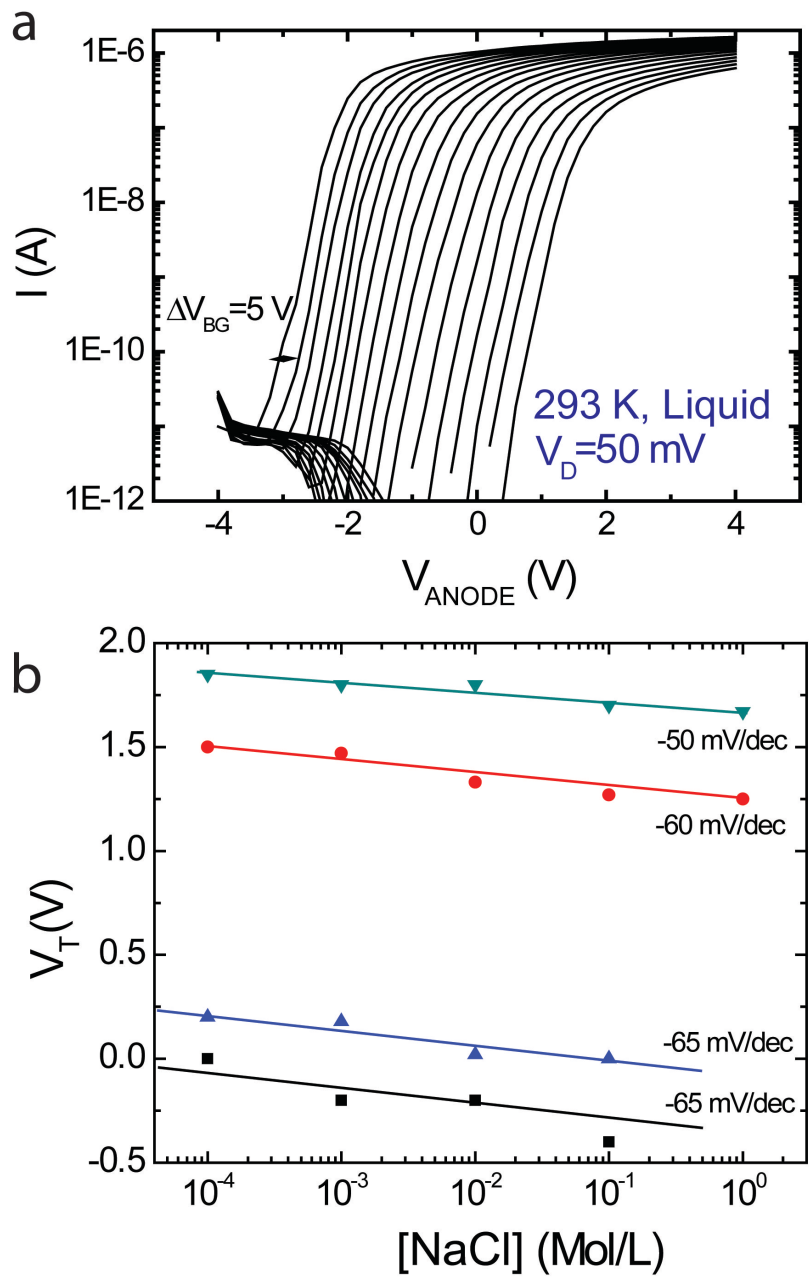

Fig.2

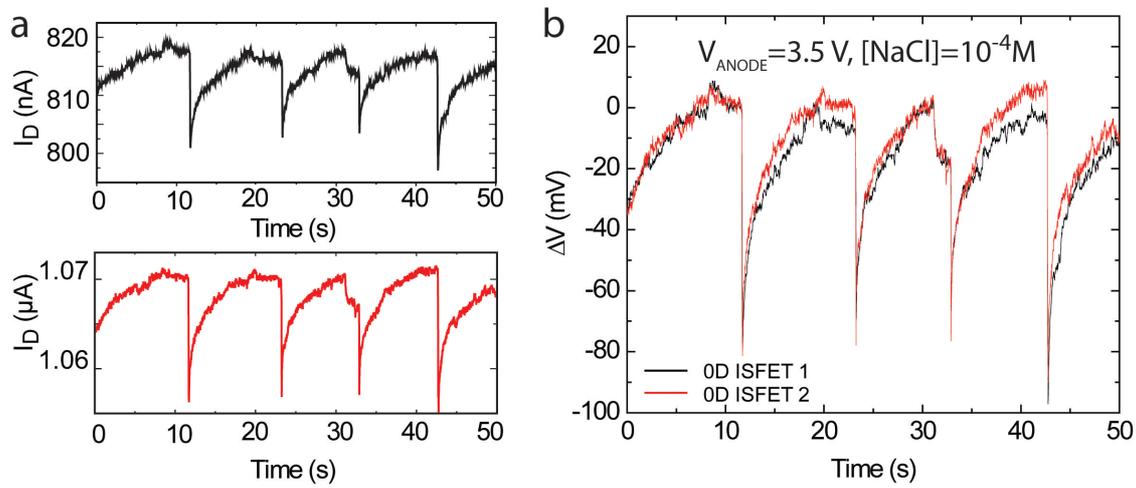

Fig.3

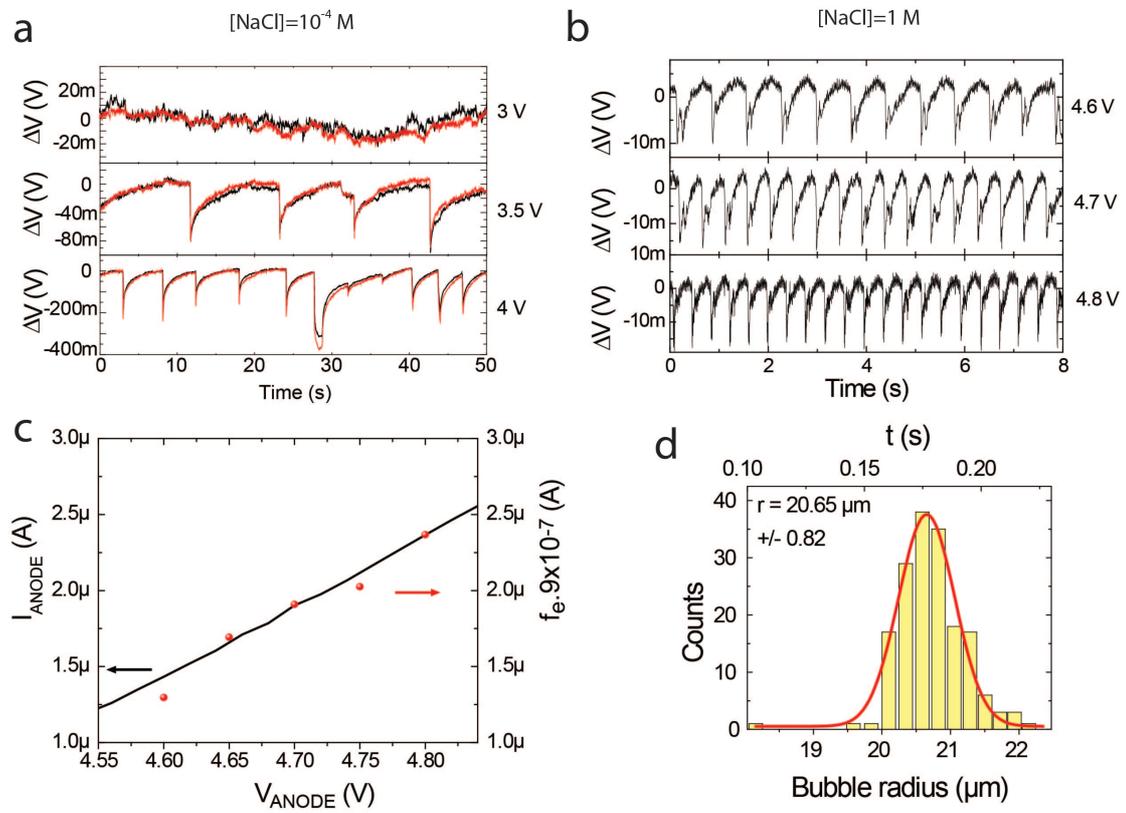

Fig.4

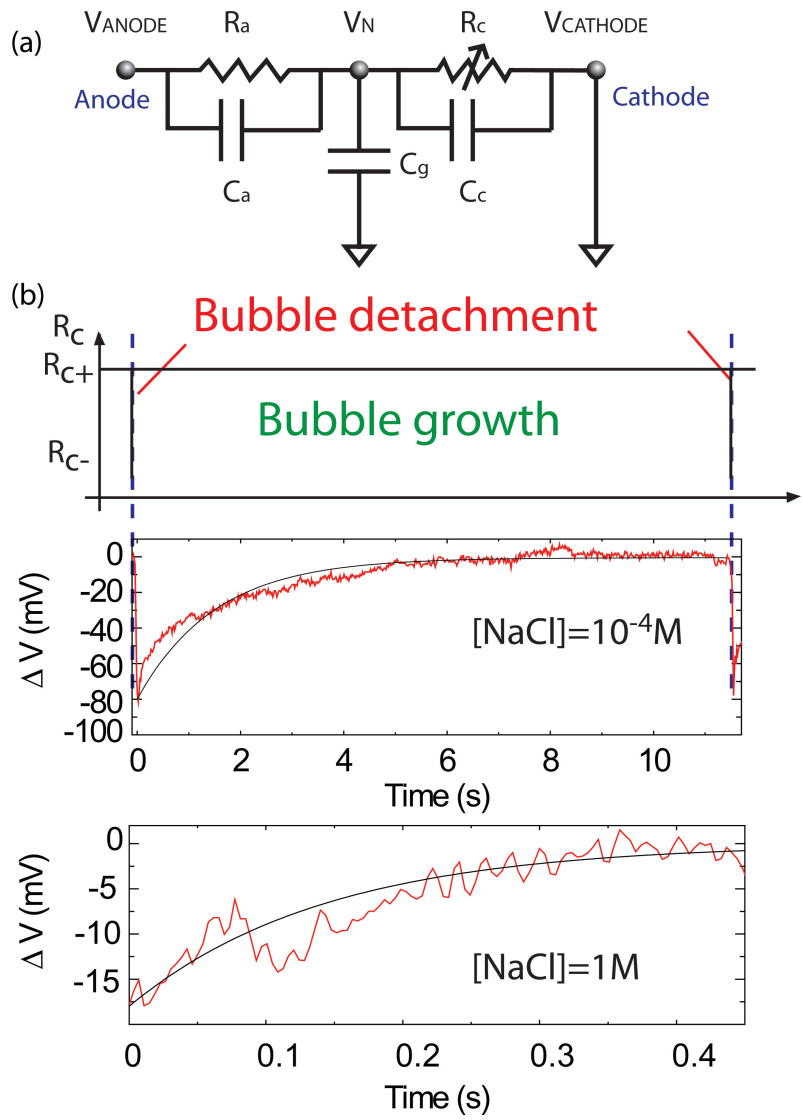

Fig.5

# Supplementary Information: Water Electrolysis and Energy Harvesting with 0D Ion-Sensitive Field-Effect Transistors

N. Clément[1], K. Nishiguchi[2], J.F. Dufreche[3], D. Guerin[1], A. Fujiwara[2], & D. Vuillaume[1]

(1) Institute of Electronics, Microelectronics and Nanotechnology, CNRS, Avenue Poincaré, 59652, Villeneuve d'Ascq France

(2) NTT Basic Research Laboratories, 3-1, Morinosato Wakamiyia, Atsugi-shi, 243-0198 Japan

(3) Université de Montpellier 2, Institute of separation chemistry Marcoule (ICSM UMR CEA/CNRS/UM2/ENSCM), 30207 Bagnols sur Cèze cedex France

*Device fabrication.*

*1- Nanotransistor fabrication:*

The nanoscale MOSFETs were fabricated on an SOI wafer. First, a narrow constriction sandwiched between two wider (400-nm wide) channels was patterned on the 30-nm-thick top silicon layer (p-type, boron concentration of $10^{15}$ cm$^{-3}$). The length and width of the constriction channel were 35 and 25 nm, respectively (Fig. 1-d). The patterning was followed by thermal oxidation at 1000 °C to form a 40-nm-thick $SiO_2$ layer around the channel. This oxidation process reduced the size of the constriction to about 15 nm, giving a final channel dimension of 20×25 nm. Then, we implanted phosphorous ions outside the constriction, 5 µm away from it, using a resist mask to form highly doped source and drain regions. Finally, aluminium electrodes were evaporated on the source and drain regions.

*2- Fabrication of the Pt microelectrode and SU8 micro-bath:*

Pt microelectrodes (100 nm) are made with conventional optical lithography techniques. The fabrication of the 300 µm – thick micro-bath is more complex. We use two layers of SU8 resist (SU8 2002, 2 µm and SU8 2075, 500 µm). The first layer is used to get a good adhesion to the substrate. Annealing should be made with a temperature ramp to avoid only surface baking. Then optical alignment process should be made in proximity mode because at this stage resist is not completely hard. After development, we make a hard baking at 190°C for two hours to polymerize the resist. An example of empty micro-bath is shown in Fig. S1a and filed in Fig.S1b.

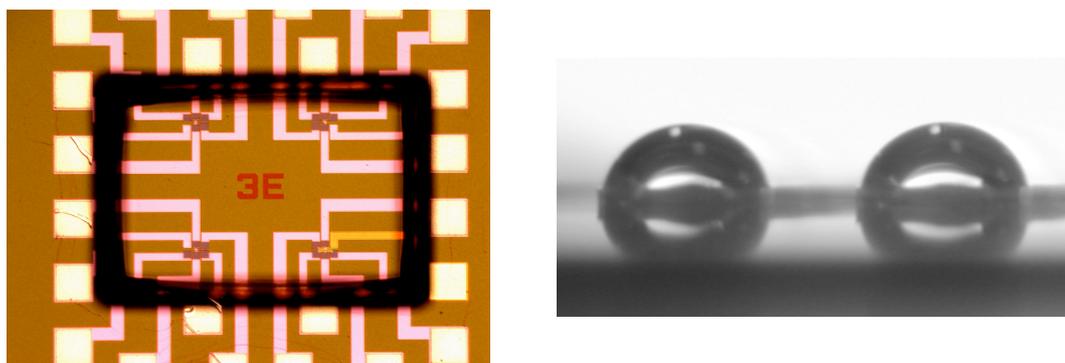

Fig.S1 (a) Optical image of 300 µm-thick micro-bath made with SU8 resist. (b) Side photography of micro-droplets inserted in micro-baths.

*3- Calibration with an Ag/AgCl reference electrode*

We have compared the droplet potential applied by the Pt electrode with an Ag/AgCl reference electrode. A schematic view of the experimental setup is shown in Fig. S2a. It is the same as the one in Fig. 1a, except the absence of cathode and the electrode added to measure the droplet voltage Vm between the droplet and the ground with Ag/AgCl potential reference. We find that as we linearly increase Vg (Platinum electrode), Vm follows Vg with an initial offset of 0.503 V (Fig.S2c).

This value is given as information, but in the article, we refer to the potential applied on the Platinum electrode. The home-made reference micro-electrode has been fabricated following a protocol described in ref 25 . An optical microscope image of the reference electrode is shown in Fig.S2b.

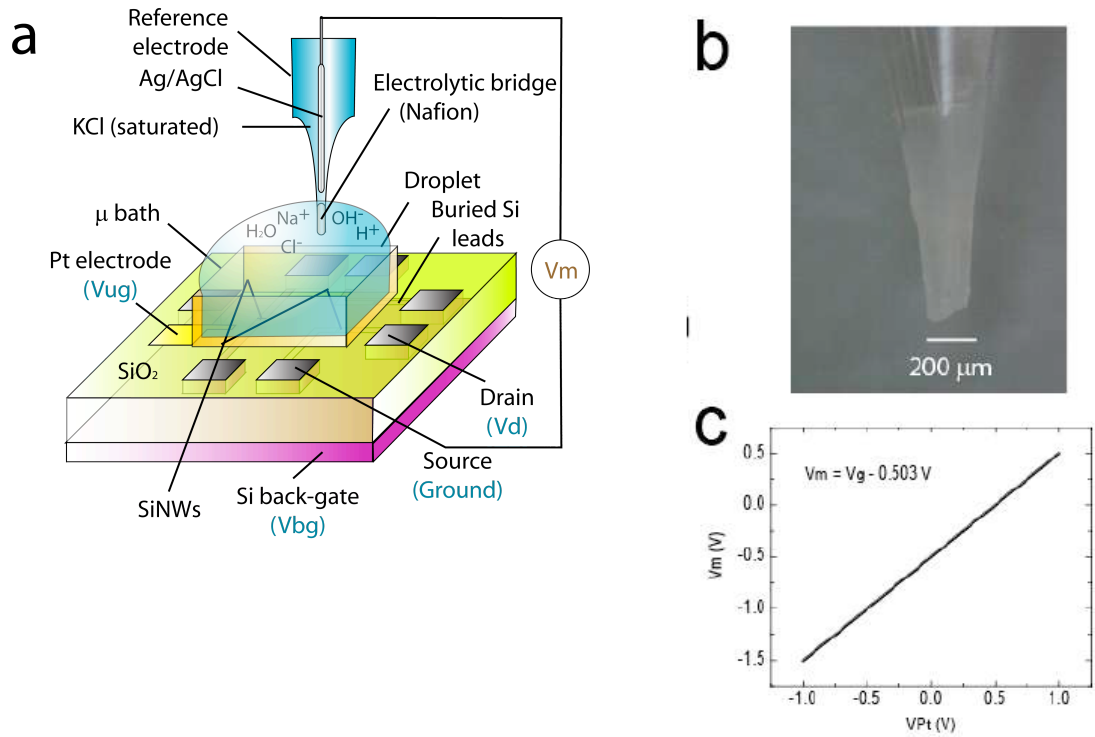

Fig.S2: Schematic view of the ISFET structure with an Ag/AgCl reference electrode (a), optical image of the reference electrode (b) and Vm measured as Vg is swept from -1 to +1 V (c).

**At equilibrium, 0D ISFETs have elementary charge sensitivity**

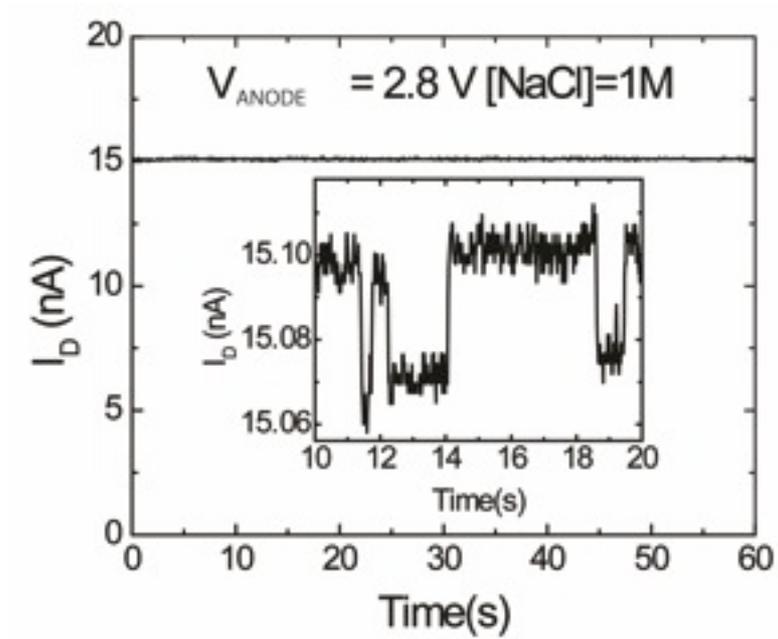

Fig.S3: Time-dependent of drain current at a given $V_{ANODE}$ bias of 2.8 V. The charge noise is so low that fluctuation of a single electron is clearly obesrved as a discrete step in the 0D ISFET (zoom: inset).

## $g_m$ extraction from $I_D$-$V_{ANODE}$ curves

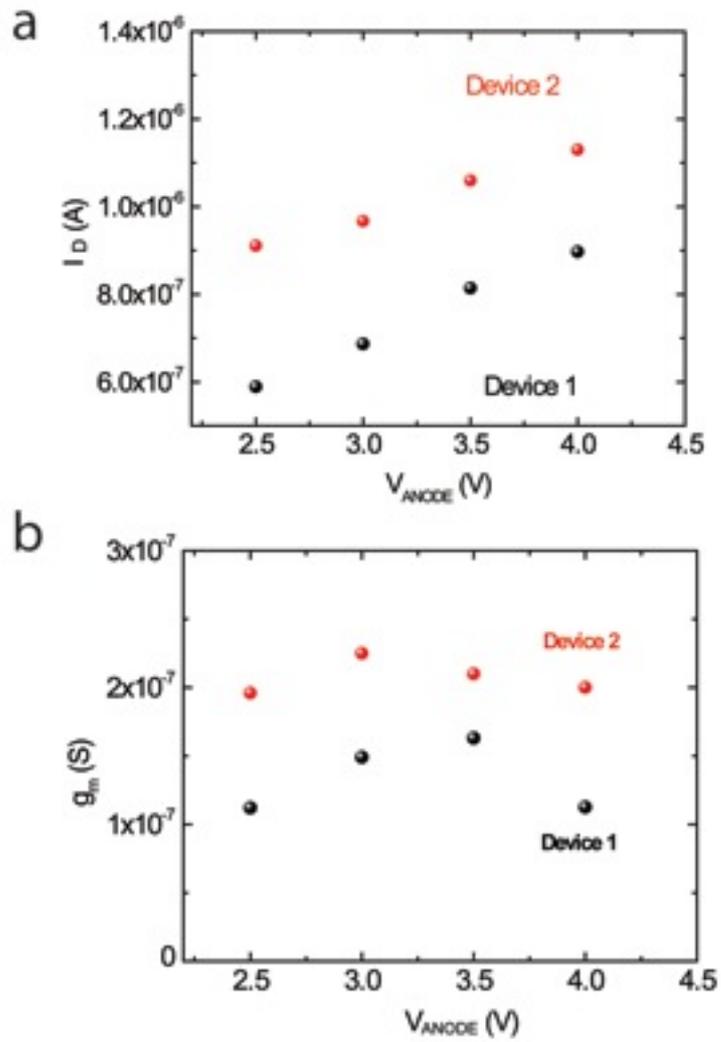

Fig.S4: $I_D$-$V_{ANODE}$ (a) and $g_m$-$V_{ANODE}$ (b) obtained by derivation of (a)

**FFTs of drain current and of anodic current**

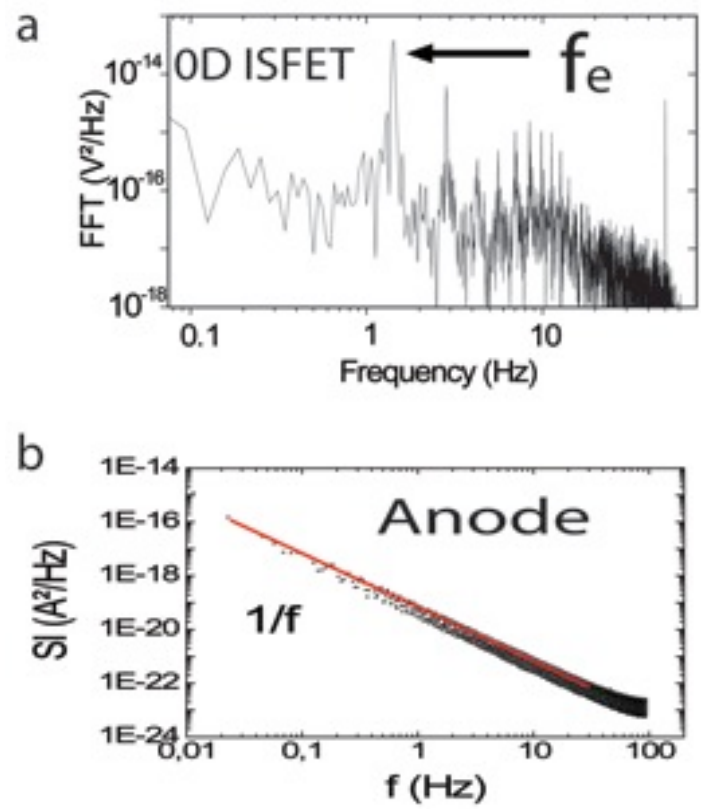

Fig.S5: Power spectrum current noise of 0D ISFET drain current (a) and anode current (b). The 1st pic in FFT (a) is used to evaluate the bubble emission frequency $f_e$.

**Bubble size estimation from optical microscope image.**

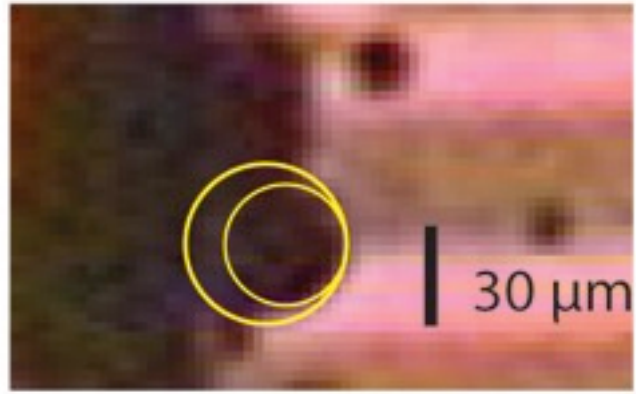

Fig.S6: Zoom on the optical image shown in Fig.1-c. Estimated bubble radius (just after detachement) is in the range 18-24 µm. Due to the limit in resolution of our camera, we indicate with yellow circles the minimum and maximum bubble size estimated.

**Measurements in 0D ISFETs when Vanode=6V (many bubbles)**

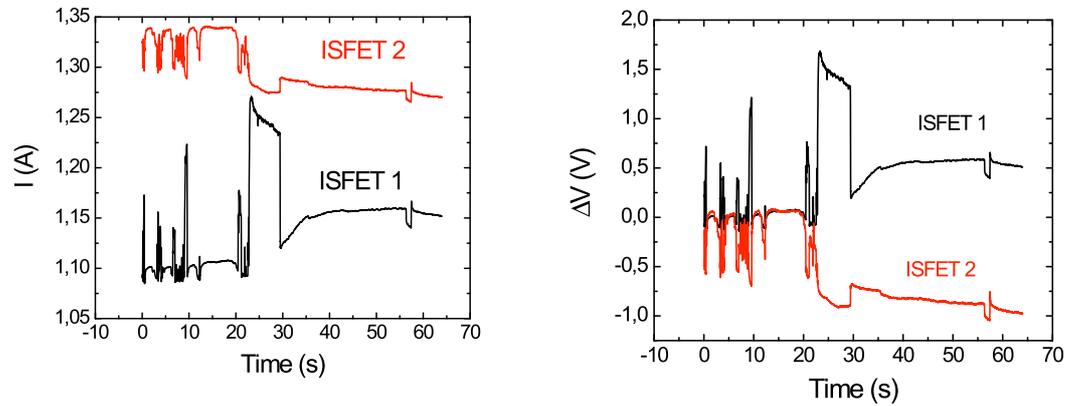

Fig.S7 I-t (left) and equivalent $\Delta$V-t (right) measurements with 2 0D ISFETs at large $V_{anode}$ (6V), corresponding to many bubbles emission. The resulting signal loses its regularity.

**Effect of dilution of OH$^-$**

For every bubble whose lifetime is 1s, we typically have a dilution in a volume V equal to 500 µm x (100 µm)². 500 µm represents the distance in the direction of electrodes reached by migration and 100 µm is the diffusion distance for 1s. The resulting concentration is 2x10$^{-3}$ M/L. This value is high compared to 10$^{-4}$ M/L (Fig.3) but relatively low compared to 1 M/L (Fig.4). Consequently, high ionic strength and ion concentrations are significantly modified in the first case but the effect is much weaker in the second one.

---

[25] T. Kitade, K. Kitamura, S. Takegami, Y. Miyata, N. Nagatono, T. Sakaguchi and M. Furukawa, Anal.Sci. 21, 907 (2005)